\documentclass[12pt]{iopart}
\usepackage{graphicx}
\usepackage{iopams}
\begin{document}
\title[Analogues of magnetic monopoles in semiconductor microcavities]{Separation and acceleration of analogues of magnetic monopoles in semiconductor microcavities}

\author{H. Flayac, D. Solnyshkov, G. Malpuech}

\address{Institut Pascal, PHOTON-N2, Clermont Universit\'{e}, Blaise Pascal University, CNRS, 24 avenue des Landais, 63177 Aubi\`{e}re Cedex, France}

\ead{hugo.flayac@univ-bpclermont.fr}

\begin{abstract}
Half-integer topological defects in polariton condensates can be regarded as magnetic charges, with respect to built-in effective magnetic fields present in microcavities. We show how an integer topological defect can be separated into a pair of half-integer ones, paving the way towards flows of magnetic charges: spin currents or magnetricity. We discuss the corresponding experimental implementation within microwires (with half-solitons) and planar microcavities (with half-vortices).
\end{abstract}
\pacs{71.36.+c,71.35.Lk}
\submitto{\NJP}
\maketitle
\section{Introduction}
Abstraction is a common tool in physics, serving the better comprehension of complex phenomena. The most evident example is the atom, encompassing an intricate underlying structure of electrons and hadrons, the latter themselves composed of quarks and gluons. This internal structure can of course be completely neglected while discussing the properties of gases in statistical physics. Such  multi-level abstraction is especially common for solid-state physics, where excitons are formed from electrons and holes, themselves being complicated elementary excitations formed from several electronic levels of atoms constituting the solid. The excitons can be described as massive quantum particles, which can couple with photons and form new particles of an even higher level of abstraction - composite bosons called exciton-polaritons.

When a Bose condensate \cite{BEC} is formed, one can consider its weak excitations as elementary particles (called Bogolons), forgetting the nature of the underlying bosons. But the weak elementary excitations are not the only type of interesting perturbations which can occur in a Bose condensate. The topological defects \cite{EmergentNonlinear} resulting from the non-linearity of the Gross-Pitaevskii equation and the irrotationality of the macroscopic condensate wavefunction are currently (and since quite a long time) in the focus of intense theoretical and experimental research and will be the main topic of the present work. We shall see how their behavior can be described in terms of relativistic "material points" and "point charges" (an easy way) or in terms of underlying local spin dynamics (a harder way). The importance of such analogies as "magnetic charges" will thus become especially clear.

In a recent paper \cite{SolnyshkovMonopoles}, we have discussed the behavior of single half-solitons\cite{VolovikHS} and half-vortices\cite{RuboHV} in spinor polariton condensates in the presence of an effective magnetic field (induced by the energy splitting between polarizations in wire and planar cavities). We have shown that such topological defects behave as "magnetic charges", accelerating along the applied magnetic field: that is, like analogues of magnetic monopoles. Such a behavior is a special feature of half-integer topological defects, which also applies to the so-called oblique half-solitons \cite{FlayacHS}. We have shown as well, that their stability is maintained thanks to the spin anisotropy\cite{ReviewSpin} of polariton-polariton interactions.

In the present work we analyze mode in details the behavior of \emph{integer} topological defects in the presence of an in-plane effective magnetic field. We show that they become unstable and split into half-integer ones, which will separate and accelerate in real space due to the interaction with the field. We concentrate on the practical ways of their experimental excitation in wire and planar microcavities in a controlled manner, in order to study their eventual separation and we discuss the combined extrinsic and intrinsic mechanisms that favor the separation.

\section{Effective magnetic fields in the Gross-Pitaevskii equation}
While the topological defects in Bose-Einstein condensates (single or multicomponent) are relatively well-known, their spin texture and their interaction with magnetic field have been much less studied (see ref.\cite{VortexReview}). First of all, let us see how an intrinsic effective magnetic field appears in the spinor Gross-Pitaevskii equation:
\begin{eqnarray}
i\hbar \frac{{\partial {\psi _ + }}}{{\partial t}} =  - \frac{{{\hbar ^2}}}{{2{m^*}}}\Delta {\psi _ + } + {\alpha _1}{\left| {{\psi _ + }} \right|^2}{\psi _ + } + {\alpha _2}{\left| {{\psi _ - }} \right|^2}{\psi _ + }\\
i\hbar \frac{{\partial {\psi _ - }}}{{\partial t}} =  - \frac{{{\hbar ^2}}}{{2{m^*}}}\Delta {\psi _ - } + {\alpha _1}{\left| {{\psi _ - }} \right|^2}{\psi _ - } + {\alpha _2}{\left| {{\psi _ + }} \right|^2}{\psi _ - }
\end{eqnarray}
Here $\alpha_1$ is the interaction constant in the triplet configuration (parallel spins), and $\alpha_2$ is the interaction constant in the singlet configuration (opposite spins), usually negative \cite{Renucci2005} and much weaker than the former due to the fact that it is a second-order process involving dark exciton states. These constants are usually calculated using the formula\cite{Tassone} $\alpha_1=6 x E_{b} a_{B}^2/S$, where $S$ is the normalization surface, $a_B$ is the exciton Bohr radius, $x$ is the excitonic fraction of the polariton state, and $E_b$ is the exciton binding energy, while $\alpha_2$ is usually assumed to be $\alpha_2=-(0.01.. 0.2) \alpha_2$, or simply neglected. Below, we are going to check the influence of this choice on the behavior of the condensates.

The spinor Gross-Pitaevskii equation can be rewritten in the following form:
\begin{eqnarray}
\nonumber i\hbar \frac{{\partial {\psi _+ }}}{{\partial t}} &=&  - \frac{{{\hbar ^2}}}{{2 m^*}}{\Delta\psi _+ } + \frac{\alpha_1+\alpha_2}{2}\left(\left|\psi_{+}\right|^2+\left|\psi_{-}\right|^2\right)\psi_{+}\\
&+&\frac{\alpha_1-\alpha_2}{2}\left(\left|\psi_{+}\right|^2-\left|\psi_{-}\right|^2\right)\psi_{+}\\
\nonumber i\hbar \frac{{\partial {\psi _- }}}{{\partial t}} &=&  - \frac{{{\hbar ^2}}}{{2 m^*}}{\Delta\psi _- } + \frac{\alpha_1+\alpha_2}{2}\left(\left|\psi_{+}\right|^2+\left|\psi_{-}\right|^2\right)\psi_{-}\\
&-&\frac{\alpha_1-\alpha_2}{2}\left(\left|\psi_{+}\right|^2-\left|\psi_{-}\right|^2\right)\psi_{-}
\end{eqnarray}
It is easily noticed that the last term enters the two equations with opposite signs, and therefore can be described by the Pauli matrix $\sigma_z$ in the Hamiltonian. It can therefore be considered as an effective magnetic field along the $z$ direction. This field is responsible for the so-called self-induced Larmor precession of the polariton pseudospin \cite{Larmor}. It is very important to note here, that the direction of this field is opposite to the $z$-projection of the polariton pseudospin. Therefore, if the condensate is completely circularly polarized with $\psi_{+}=\sqrt{n}$, $\psi_{-}=0$, $n$ being the total density, the effective magnetic field will point in the \emph{negative} direction of the $z$-axis:
\begin{equation}\label{Omegaz}
    \Omega_{z}=-\frac{\alpha_1-\alpha_2}{2\hbar}\left(\left|\psi_{+}\right|^2-\left|\psi_{-}\right|^2\right)
\end{equation}
In the polaritons case, where $\alpha_2$ is around ten times weaker than $\alpha_1$ but negative, the effective field is strengthened and tends to lock circularly polarized states (strong density imbalance) in the system. It provides the natural stability of half-integer topological defects that carry particles with spin up ($\sigma_+$) or down ($\sigma_-$) at their core. In usual spin isotropic atomic condensates $\alpha_1\simeq\alpha_2$, and therefore $\Omega_{z}\simeq0$.

Another effect, which is well known in planar or wire microcavities, is the splitting between the orthogonal linear polarizations at $\vec{k}=\vec{0}$ (as opposed to the wavevector-dependent TE-TM splitting). This splitting is especially well resolved above the condensation threshold \cite{Kasprzak2006} thanks to the sharpening of the emission lines and induces an in-plane effective magnetic field pointing in a well defined direction, determined by the orientation of the cristallographic axes of a planar cavity or the orientation of the microwire\cite{Dasbach2005}. Its interpretation in terms of an constant effective in-plane magnetic field is even more direct than in the case of the interaction-induced field in the $z$ direction: if the $x$ and $y$ polarizations have different energies when the Gross-Pitaevskii equation is written on the $xy$ basis, this splitting transforms into a term $-\hbar\Omega_{LT}\psi_{\mp}/2$ in the circular polarization basis with the usual coordinate transformation rules ${\psi _ \pm } = \left( {{\psi _x} \mp i{\psi _y}} \right)/\sqrt 2$.

For a homogeneous wavefunction, the kinetic energy term is absent, and one can easily see that the spinor Gross-Pitaevskii equation for the wavefunction can be converted into an equation for the dynamics of the pseudospin vector $\vec{S}$ with the components given by:
\begin{eqnarray}
\nonumber {S_x} &=& {\mathop{\Re}\nolimits} \left( \psi _{+} \psi_-^*\right)\\
{S_y} &=& {\mathop{\Im}\nolimits} \left( \psi _{-} \psi_+^*\right)\\
\nonumber {S_z} &=& \left({{\left| {{\psi _ + }} \right|^2} } - {{\left| {{\psi _ - }} \right|^2}}\right)/2
\end{eqnarray}
The equation for the pseudospin dynamics is very simple. It is obtained from the equation for the wavefunction by multiplying the corresponding equations by the wavefunction or its conjugate and taking the real or imaginary parts, in order to obtain the $x$ and $y$ projections. In fact, it is the usual equation for the precession of a magnetic momentum vector in a magnetic field:
\begin{equation}
\frac{\partial\vec{S}}{\partial t} = \vec{\Omega}  \times \vec{S}
\end{equation}
However, the field itself depends on the magnetic momentum via the non-linear term
\begin{equation}\label{Omega}
    \vec{\Omega}=\Omega_{LT}\vec{u}_{x}-\frac{\alpha_{1}-\alpha_{2}}{2\hbar} S_z\vec{u}_{z}
\end{equation}

\section{Spin-topological defects in a magnetic field}
The scalar 1D Gross-Pitaevskii equation is known to possess grey soliton solutions \cite{BEC}
\begin{equation}
\label{soliton}
\psi\left(x-vt\right)=\sqrt{n}\left[i\frac{v}{c}+\sqrt{1-\frac{v^2}{c^2}}\tanh\left(\frac{x-vt}{\xi\sqrt{2}}\sqrt{1-\frac{v^2}{c^2}}\right)\right]
\end{equation}
here $v$ is the speed of the soliton, $\xi=\hbar/\sqrt{2m\alpha n}$ is the healing length, and $c=\sqrt{\alpha n/m}$ is the speed of sound. Let us consider the time evolution of the simplest extension of this scalar solution to a vectorial condensate: a half-soliton. In this part of the text, we will neglect the small interaction between opposite spins $\alpha_2$ for the sake of simplicity. The \emph{dark} half-soliton solution for which $v=0$ reads \cite{FlayacHS}:
\begin{eqnarray}
\psi_{+}=\sqrt{{\frac{n}{2}}}\tanh\left(\frac{x}{\xi\sqrt{2}}\right),  \psi_{-}=\sqrt{{\frac{n}{2}}}
\end{eqnarray}
where we have assumed no phase difference between the two components. This solution is shown in the figure 1(a) together with its pseudospin pattern. One can note that the half-soliton is a domain wall between the two regions of perpendicular linear polarizations (opposite pseudospins). We remind that the in-plane pseudospin field $\vec{S}_{\parallel}=(S_x,S_y)^T$ describing linear polarized states, makes a double angle with respect to the polarization angle. We see that the pseudospin pattern is divergent in that case, resembling the field of a point (magnetic) charge. The sign of the charge is given by the pseudospin pattern of the half-soliton (which is not necessarily divergent) and is defined by the relative phase between the two components.
\begin{figure}[ht]
\includegraphics[width=0.8\linewidth]{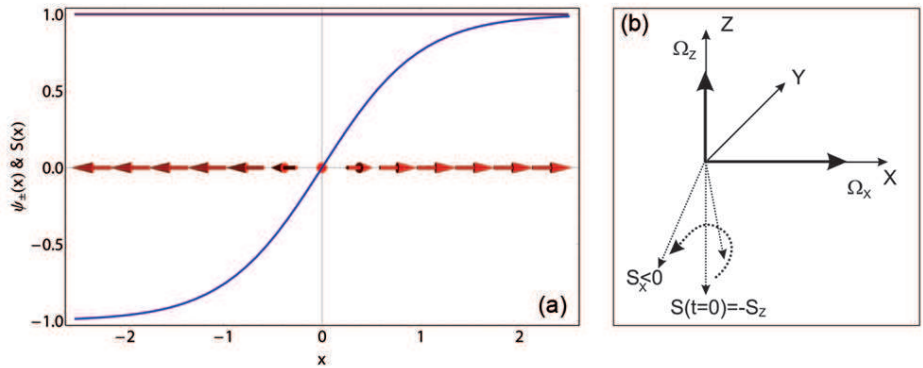}
\centering
\caption{(a) The two circular components $\sigma_+$ (blue line) and $\sigma_-$ (purple line) of the half-soliton wavefunction with the corresponding pseudospin vector field (red arrows). (b) Precession of the pseudospin under the action of the effective field $\vec{\Omega}$.}
\label{fig1}
\end{figure}
Let us first try to understand from "microscopic" considerations what will happen to this object when an in-plane effective magnetic field $\vec{\Omega}_{LT}=\Omega_x\vec{u}_x$ is applied. For this, one can consider the pseudospin dynamics at each point, since the two other terms in the Gross-Pitaevskii equation (the kinetic energy and the interaction energy described by the blue shift) compensate each other at $t=0$. For the regions far from the soliton core $|x|\gg\xi$, the pseudospin is aligned or anti-aligned with the magnetic field (which contains only a $x$ component) and there is no evolution: $\partial \vec{S}/\partial t=\vec{0}$. For the core region, there is a non-zero pseudospin projection on the $z$ axis, and therefore the pseudospin will rotate (precess) around the magnetic field (which now contains both $x$ and $z$ components). Let us consider the initial moments of this rotation for the pseudospin in the center of the soliton: $\vec{S}(x=0,t=0)=\vec{u}_z$, while the total magnetic field is
\begin{equation}\label{Omega0}
\vec{\Omega}(x=0,t=0)=\Omega_{x}\vec{u}_{x}-\frac{\alpha_{1}-\alpha_{2}}{2\hbar}(-\frac{n}{2})\vec{u}_{z},
\end{equation}
Rotating around the (positive) $\Omega_x$ component, the pseudospin (initially negative along $z$) gains a positive $S_{y}$ projection and starts to precess around the positive $\Omega_z$ field, turning towards the negative direction of the $x$-axis [see Fig.1(b)]. This is the main result of our qualitative vectorial consideration of the polarization dynamics: the pseudospin in the center of the soliton gains a {\emph{negative}} $x$-projection. Therefore, the domain of negative $x$ pseudospin projection becomes larger, the domain of positive $x$-projection smaller, and the wall between these domains is moving to the right. However, once the soliton core starts to propagate, the kinetic and interaction energy terms are no more compensated everywhere, and one cannot discuss the evolution of the system using the qualitative arguments based on polarization dynamics. At first sight, one may even think that the nonlinear system in question can be solved only numerically.

However, an important insight into the behavior of the system can be gained by "changing the zoom". Forgetting about the complicated internal structure, a vectorial grey soliton in a Bose-Einstein condensate can be considered as a particle\cite{VolovikReview} with a negative effective mass (at least at low velocities). Moreover, the pseudospin pattern of this particle is the same as the field of a point magnetic charge in 1D. The magnetic energy of the system can be found from the Hamiltonian as the usual scalar product of the field and the spin, and this magnetic energy depends on the position of the soliton because of the finite system size. Thus, one can evaluate the force acting on the magnetic charge from the magnetic field as a gradient of the magnetic energy with respect to the position of the soliton. This force will consequently accelerate the soliton.

Considering the soliton as an elementary particle without internal structure means passing to the limit $L\gg\xi$, where $L$ is the system size (for example, the length of the wire cavity, which is usually of the order of 100$\mu$m). The healing length of a polariton condensate for a reasonable blue shift of 1 meV expected for GaAs or CdTe cavities and a polariton mass of $5\times 10^{-5}$ of a free electron mass is $\xi=0.8 \mu$m.
In this limit, the wavefunction of the soliton at $t=0$ becomes simply $\psi_{+}=\sqrt{n/2}{\rm sign}(x-x_0)$, $\psi_{-}=\sqrt{n/2}$, where $x_0$ is the position of the soliton. In general, the \textit{tanh} function is replaced by the \textit{sign} function. The magnetic energy of a condensate containing a HS in an external in-plane magnetic field is
\begin{equation}\label{Emag}
E_{mag}=-\int (\vec{\Omega}/2\cdot\vec{S}) dx
\end{equation}
Here $\vec{S}(x-x_{0})=n {\rm sign}(x-x_{0})\vec{u}_x$ ($x_0$ is the soliton position) in the limit we consider, giving
\begin{eqnarray}
{E_{mag}}\left( {x - {x_0}} \right) =   \hbar {\Omega _x}n\left( {x - {x_0}} \right)\\
{F_{mag}} =  - \frac{{d{E_{mag}}}}{{d{x_0}}}=-n \Omega_x
\end{eqnarray}
The force in Eq.(14) is therefore acting opposite to the direction of the magnetic field, but the acceleration will occur in the direction opposite to the force, because the mass of the soliton (which is a density notch in the condensate) is negative, at least at low velocities.

For a grey soliton propagating at speed $v$, the phase shift induced by the soliton in the $\sigma_{+}$ component is $\Delta \theta=2 {\rm arccos}(v/c)<\pi$ and the pseudospin projection $S_x$ is reduced, which can be expressed as a renormalization of the magnetic charge. The correction to the charge is found as $q=q_0(1-v^2/c^2)$ ($q_0$ is the charge at rest for a dark half-soliton) by integrating the solution (\ref{soliton}) in the limit $L\gg\xi$. The total correction for the mass of the soliton and its charge gives the equation of motion
\begin{equation}\label{Acceleration}
a = {q_0}\frac{{n{\Omega _x}}}{{m_0}}{\left( {1 - \frac{{{v^2}}}{{{c^2}}}} \right)^{3/2}}
\end{equation}
the same as in relativistic physics, integrating which one obtains
\begin{equation}\label{Velocity}
v(t)=c\tanh \left( {\frac{{{q_0}{\Omega _x}n}}{c}t} \right),
\end{equation}
assuming zero initial velocity. This trajectory is perfectly confirmed by numeric simulations as shown in Ref.\cite{SolnyshkovMonopoles}.

Thus, using the abstraction of a point magnetic charge proves particularly useful, because it allows solving the nonlinear spinor Gross-Pitaevskii equation analytically and gives a good qualitative understanding of observed phenomena. In the following we will discuss the soliton interaction and how to realistically generate currents of half-solitons and half-vortices within semiconductor microcavities toward "polariton magnetricity", similarly to the "magnetricity" reported in spin ices \cite{Magnetricity2009}, but with much higher flow velocities (currents).

\subsection{Soliton and half-solitons interactions}
\begin{figure}[ht]
\includegraphics[width=0.85\linewidth]{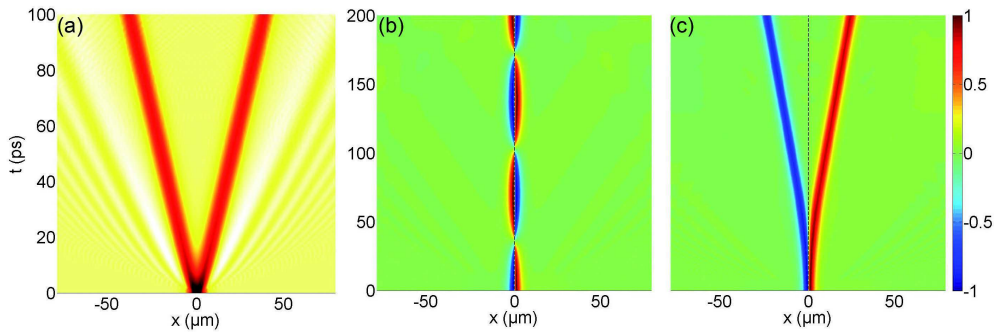}
\centering
\caption{Short range interaction between solitons initially separated by 0.5 $\mu$m. (a) Integer soliton repulsion the colormap shows the density $n={\left| \psi  \right|^2}$ of the single component of a scalar condensate. (b) Half-solitons attraction, which shows dipolar oscillations for $\alpha_2=+0.2\alpha_1$ (c) half-solitons repulsion for $\alpha_2=-0.2\alpha_1$, the colormap in (b) and (c) shows the degree of circular polarization of the spinor condensate: $\rho_c=(n_+-n_-)/(n_++n_-)$.}
\label{fig2}
\end{figure}

Dark (grey) solitons are the solution of the 1D Gross-Pitaevskii equation provided that the condensate is formed of particles that repel each other ($\alpha_1>0$). Interactions between solitons themselves have been thoroughly analyzed an it is now well known that dark solitons tend to repel each other as well \cite{SolInteractions}. Such a behavior is illustrated in the Fig.\ref{fig2} showing the degree of circular polarization and where we have solved the scalar Gross-Pitaevskii equation with the initial condition given by two dark solitons spatially shifted by $0.5$ $\mu$m [see Eq.(\ref{HSPair}) below]. The solitons repel at the initial moment and then demonstrate a linear trajectory, which is a clear signature of their \emph{short range} interaction.

Now, what about the half-soliton in two-component condensates? Let us remind that a half-soliton corresponds to a soliton occurring in only one of the two components. Additionally, if the intercomponent interaction is not neglected, the presence of a half-soliton in one component obviously perturbs the other one in a fashion depending on the type of interactions, given by the sign of $\alpha_2$. Indeed, for $\alpha_2>0$ ($\alpha_2<0$), the presence of a density minimum in the component carrying the soliton is seen as a potential well (barrier) by the other one and consequently a density maximum (minimum) appears at the position of the soliton (this effect is discussed in details with an analytical estimate of the density perturbation in Ref.\cite{FlayacHS}). Putting two half-solitons lying in  different components next to each other induces a short range interaction within each component, between a soliton and the density maximum (minimum) induced indirectly by the other soliton. The half-soliton having a negative effective mass are repelled by the potential wells they encounter (which, from their point of view, are barriers). This argumentation leads to the following statement: For $\alpha_2>0$ ($\alpha_2<0$), half-soliton attract (repel) each other while they obviously don't interact for $\alpha_2=0$. This trend is shown in Fig.2(b) and (c). We note, that in the repulsive case, dipolar oscillations occur between the solitons, this is due to the effect of their mutual attraction and the slight initial displacement (the system can be described by a simple pendulum in this case). In the framework of polariton condensates where $\alpha_2<0$, the repulsion of half-solitons is therefore a strong asset for the separation of integer soliton into a pair of half-solitons of opposite charges.

We note that bright solitons have also been described for exciton-polaritons \cite{Skryabin1, Skryabin2}, but these are not in the focus of the present work, although the same considerations of the interactions with magnetic field would apply to them as well.

\subsection{Solitons imprinting using Gauss-Laguerre beams}
In a recent work \cite{SolnyshkovBH} we have made a proposal for the creation of half-integer (valid also for integer) solitons in quasi-1D polariton condensates under continuous  quasi-resonant pumping. It consists in applying pulsed potential either equally to both circular components (integer topological defects) or to a single one (half-integer defects). This could be experimentally realized by sending short optical pulses at detuned frequency with respect to the main pumping laser. Such a setup would however require the use of two lasers and the $cw$-resonant injection would tend to imprint the phase of the condensate at any time under the pump spot, which would be harmful to the solitons characterized by local phase dislocations. Another method of hydrodynamic nucleation of solitons has been recently proposed theoretically in Ref.\cite{Pigeon} and then implemented experimentally\cite{AmoSolitons}. However, this method requires the use of a defect, whose properties cannot be controlled at will. Here we propose a simpler configuration for the creation of half and integer solitons in quasi-1D microwires, while the 2D case will be discussed in the next section.

Vortices have already been created in polaritons condensates by several means, among which is found the artificial phase imprinting induced by a probe carrying an angular momentum: a "Gauss-Laguerre (GL) beam" \cite{GLBeam}. Such a beam is obtained from a usual Gaussian beam scattered on a hologram containing a forklike dislocation. In numerical simulations, this type of pumping can be described by the following complex function
\begin{eqnarray}\label{GL}
\nonumber {P_{GL}}\left( {\vec{r},t,l} \right) &=& A_{GL}\sqrt {{(x-\Delta x)^2} + {(y-\Delta y)^2}} \\
&\times& {e^{ - {{\left( {\frac{{x - {\Delta x}}}{{\sigma_x}}} \right)}^2}}}{e^{ - {{\left( {\frac{{y - {\Delta y}}}{{\sigma_y}}} \right)}^2}}}{e^{ - {{\left( {\frac{{t - {\Delta t}}}{{\sigma_ t}}} \right)}^2}}}{e^{il\phi }}{e^{ - i{\omega _{GL}}t}},
\end{eqnarray}
written here in Cartesian coordinates. The exponentials give the "Gauss" part of the function, while the square root is the 1$^{st}$ order Laguerre polynomial, giving zero density in the center of the beam. $l$ is the integer winding number that is to be transferred to the vortex state. The laser frequency $\omega_{GL}$ should be slightly blue-detuned from the bare polariton mode, in order to make use of the bistability effect and obtain an almost flat density profile of the condensate except in the center, due to the saturation of the pumping efficiency on the upper bistability branch\cite{Gippius}.

\begin{figure}[ht]
\includegraphics[width=1\linewidth]{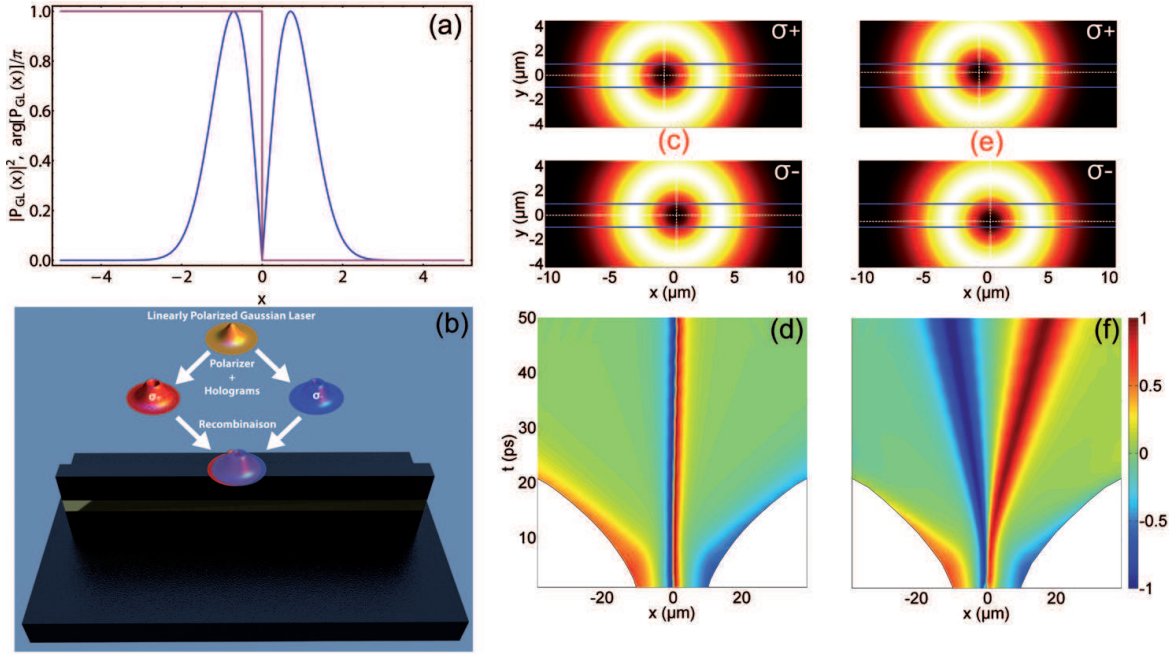}
\centering
\caption{(a) Normalized slice of a GL beam together with its phase (for the case $l=+1$) as seen by a pure 1D system. (b) Scheme of the separation of a Gaussian beam into two GL beams subsequently shined on a wire shaped microcavity. (c)-(f) Impact of the traverse displacement of the spots, here $\alpha_2=0$ and $(l_+,l_-)=(-1,-1)$. (c)-(d) No transverse displacement leads to static dark HSs in each component. (e)-(f) A small $y$ shift leads to a uniform propagation of the HS. We see that the direction and speed of propagation depends on the direction of the shift sign and on the proximity of the spot center from a boundary respectively. In (e) the $\sigma_+$ ($\sigma_-$) beam is shifted by $\Delta y_+=+0.25$ $\mu$m ($\Delta y_-=-0.50$ $\mu$m). The blue lines in (c) and (e) stand for the wire's boundaries while the white dashed lines show the position of the spot center. In (d) and (f) the white regions are absent of particles due to the finite size of the spot.}
\label{fig2bis}
\end{figure}

As one can see from the sixth factor of Eq.(\ref{GL}), where $\phi$ is the polar angle, the phase is changing continuously from $0$ to $2l\pi$ encircling the beam center. In a pure 1D system, the notion of angular momentum vanishes and grey solitons embody elementary topological excitations in BECs in place of vortices. The action of a GL beam is therefore to induce a local phase step of $-l\pi$ or smaller, equal to the one resulting from a cut of a vortex by a plane. It is illustrated in Fig.\ref{fig2bis}(a) showing a density slice of the GL beam together with its phase. This technique constitutes an efficient mean for soliton engineering in 1D condensates. We note that, in order to allow the soliton to evolve freely, it is necessary to use a pulsed GL beam, which will also form the background condensate for sufficient pump intensity.

Besides, experimental creation of half-solitons with independent selection of their phase (given by $l$) would require to separate a linearly polarized input laser in its two circularly polarized components ($\sigma_\pm$) using polarizers, to make one component (for a single half-soliton) or both of them (for a pair of half solitons) scatter on a hologram, and to recombine them on the sample. For the case where both components carry an angular momentum it looks difficult to recover a perfect spatial overlap of the two beams, which will naturally trigger the separation of the half-solitons, especially for the normal case of $\alpha_2<0$. Additionally a relative phase $\phi_0$ can be introduced between the two beams increasing the optical path of one component with respect to the other. $\phi_0$ has a crucial impact on the pseudospin textures of HSs and therefore on their interaction with effective fields as follows from Eqs.(7) and (12). A scheme of an experimental setup is shown in the Fig.\ref{fig2bis}(b).

If we consider now a more realistic system such as a microwire [see Fig.\ref{fig2bis}], we should take into account the transverse width of the sample and treat the case of a quasi-1D condensate for which the angular momentum cannot be neglected anymore. The central position of the pump spot crucially impacts the symmetry of the imprinted flow. We have performed numerical simulations of a 2 $\mu$m wide and 100 $\mu$m long wire using polariton parameters to highlight this feature. On one hand, if the spot is transversally centered, a static and therefore dark soliton is nucleated in the wire, almost as in the pure 1D case [see Fig.\ref{fig2bis}(c),(d)]. On the other hand, a shift of the GL beam along the $y$-axis induces an extrinsic uniform propagation of the resulting grey soliton in a direction depending on the transverse shift and on the sign of the imprinted winding number $l$ as it is illustrated in the figure \ref{fig2bis}(c)-(f). A positive (negative) $y$-shift induces a soliton propagation to the right (left) for $l<0$ and reciprocally for $l>0$. This effect can be qualitatively seen as a "rolling" of the particles on the boundaries, introduced by a gradient of angular velocity in the transverse direction. Moreover, the closer is the spot center to a boundary, the faster the soliton moves along the wire. This "extrinsic" motion has to be taken into account for an experiment, in which imposing a perfectly centered spot is elusive. Even in the simulation [see Fig.\ref{fig2bis}(d)], where the precision on the centering is about 0.1 $\mu$m, we observe a very small drift of the solitons.

For the simulations, we use the typical parameters of modern GaAs cavities, including their outstanding lifetimes of about 30 ps \cite{Tanese2012}. We solve the spinor Gross-Pitaevskii equations for the photonic $\phi(\vec{r},t)$ and excitonic fields $\chi(\vec{r},t)$ (coupled via the Rabi splitting $V_R=15$ meV), fully taking into account the polarization $\sigma_\pm=\pm$, the finite lifetimes $\tau_{\phi}=25$ ps and $\tau_{\chi}=300$ ps and the injection of the particles via $P_{GL}^\pm(\vec{r},t,l_\pm,\phi_0)$:
\begin{eqnarray}
i\hbar \frac{{\partial {\phi _ \pm }}}{{\partial t}} &=&  - \frac{{{\hbar ^2}}}{{2{m_{\phi}}}}\Delta {\phi _ \pm } + \frac{{{V_R}}}{2}{\chi _ \pm } - \frac{{i\hbar }}{{2{\tau _{\phi}}}}{\phi _ \pm } + {P_{GL}^\pm} + U{\phi _ \pm } - \frac{{\hbar {\Omega _x}}}{2}{\phi _ \mp }\\
i\hbar \frac{{\partial {\chi _ \pm }}}{{\partial t}} &=&  - \frac{{{\hbar ^2}}}{{2{m_{\chi}}}}\Delta {\chi _ \pm } + \frac{{{V_R}}}{2}{\phi _ \pm } - \frac{{i\hbar }}{{2{\tau _{\chi}}}}{\chi _ \pm } + \left( {{\alpha _1}{{\left| {{\chi _ \pm }} \right|}^2} + {\alpha _2}{{\left| {{\chi _ \mp }} \right|}^2}} \right){\chi _ \pm }
\end{eqnarray}
Here $m_{\phi}=3.6\times10^{-5}m_0$, $m_{\chi}=0.4 m_0$ and $m_0$ are the cavity photon, the quantum well exciton and the free electron masses respectively. $U(\vec{r})$ is a potential (e.g. the confinement potential for quasi-1D wires, or a wedge potential, see next section).

\subsection{Natural acceleration of half-solitons in microwires}
Armed with efficient means of creating half solitons, let us now argue on their potential propagation and acceleration.
We shall consider the following situation: A pair of half-solitons is created by the separated GL beams. They are slightly spatially shifted either initially or naturally due to an intrinsic noise (checked) that breaks the symmetry between the two components. Several contributions will lead to the separation, evolution and acceleration of half-solitons. First, their dissociation is triggered by a negative value of $\alpha_2$ as discussed previously and can be emphasized by the rolling effect imposed by the GL pump. Second, after the unbinding, half-solitons start to feel the longitudinal effective field. Let us now discuss the impact on their evolution.

\begin{figure}[ht]
\includegraphics[width=0.5\linewidth]{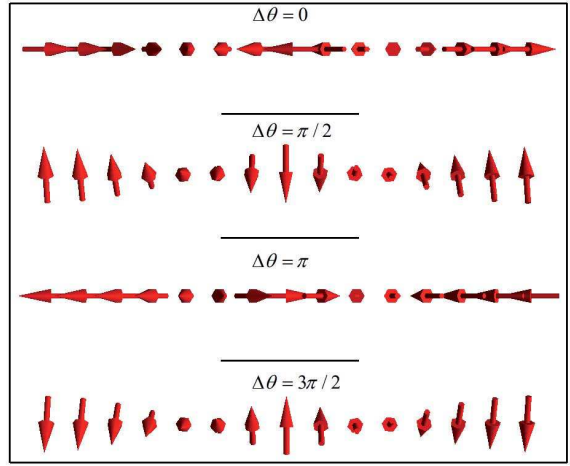}
\centering
\caption{Pseudospin textures of four separated half-soliton pairs depending on the total relative phase $\Delta\theta$.}
\label{SolTextures}
\end{figure}

The normalized condensate wavefunction carrying a dark soliton in both components reads
\begin{equation}\label{HSPair}
\vec \psi   = \left( \begin{array}{l}
{\psi _ + }\\
{\psi _ - }
\end{array} \right) = \sqrt {\frac{n}{2}} \left( \begin{array}{l}
\tanh \left[ {{l_ + }\left( {x - d/2} \right)} \right]{e^{i{\phi _0}}}\\
\tanh \left[ {{l_ - }\left( {x + d/2} \right)} \right]
\end{array} \right)
\end{equation}
assuming a variable spatial separation $d$ and $\alpha_2=0$. $\phi_0$ is a constant relative phase between the two components and $(l_+,l_-)$ can take independently the values $\pm1$ determining the sign of the $\pi$ phase shift through each soliton. The pseudospin components are straightforwardly calculated using the Eqs.(6) giving
\begin{eqnarray}
\nonumber {S_x} &=& \frac{n}{2}\cos \left( {{\phi _0}} \right)\tanh \left[ {{l_ + }\left( {x - \frac{d}{2}} \right)} \right]\tanh \left[ {{l_ - }\left( {x + \frac{d}{2}} \right)} \right]\\
{S_y} &=& \frac{n}{2}\sin \left( {{\phi _0}} \right)\tanh \left[ {{l_ + }\left( {x - \frac{d}{2}} \right)} \right]\tanh \left[ {{l_ - }\left( {x + \frac{d}{2}} \right)} \right]\\
\nonumber {S_z} &=& \frac{n}{4}\tanh {\left[ {{l_ + }\left( {x - \frac{d}{2}} \right)} \right]^2} - \frac{n}{4}\tanh {\left[ {{l_ - }\left( {x + \frac{d}{2}} \right)} \right]^2}
\end{eqnarray}
In the case where $d=0$, it is seen that the orientation of $\vec{S}$ is homogeneous and fixed by the total relative phase $\Delta\theta=\phi_0\mathrm{sign}(l_+/l_-)$. On the other hand for $d\neq0$, while the homogeneous pseudospin texture remains constant far from the solitons cores, it is of course modified between them. Comparing e.g. $\vec S_\parallel(\pm \infty)$ and $\vec S_\parallel(0)$, it is easily seen that the direction of the in-plane pseudospin is always opposite between the solitons and away from them. This is a crucial point since depending on the orientation of the effective magnetic field $\vec\Omega_{LT}$, the half-soliton separation will occur in opposite directions. For example with the set $(l_+,l_-,\phi_0)=(+1,+1,0)$, $\Delta\theta=0$ and therefore $\vec S_\parallel(\pm \infty)=+S_x \vec{u}_x$ while $\vec S_\parallel(-d/2\rightarrow d/2)=-S_x \vec{u}_x$. A field $\vec\Omega_{LT}=\Omega_x \vec{u}_x$ is parallel to $\vec S_\parallel$ far from the solitons bringing negative contribution in to $E_{mag}$ [see Eq.(\ref{Emag})], the latter is therefore maximized increasing the spacing between the solitons, where $\vec\Omega_{LT}$ and $\vec S_\parallel$ are antiparallel, the solitons are consequently accelerated in opposite directions (because their mass is negative). In this configuration, the integer soliton is unstable against the field, the slightest symmetry breaking (e.g. some noise) between the $\sigma_+$ and $\sigma_-$ components result in its decay into half-solitons. The situation is opposed for $(l_+,l_-,\phi_0)=(+1,+1,\pi)$.  We finally note that $\Delta\theta=(p+1/2)\pi$ ($p\in\mathbb{Z}$) implies $\vec S_\parallel(x)\cdot\vec\Omega_{LT}$=0 giving $E_{mag}(d)=E_0$: An integer soliton or a pair of half-solitons carry the same magnetic energy whatever $d$. However, as soon as one of the half-solitons starts to move, the $\vec S_\parallel(x)\cdot\vec\Omega_{LT}$=0 is no longer verified, and the half-solitons become accelerated by the field. Considering separately each half-solitons, their charge can be expressed as $q=q_0(1-v^2/c^2)$ with $q_0=\mathrm{sign}(l)\cos(\phi_0)$ and the divergent monopole-like texture appears e.g. for $(l,\phi_0)=(+1,0)$. We show in the Fig.\ref{SolTextures} several types of soliton textures depending on the value of $\Delta\theta$.

\begin{figure}[ht]
\includegraphics[width=1\linewidth]{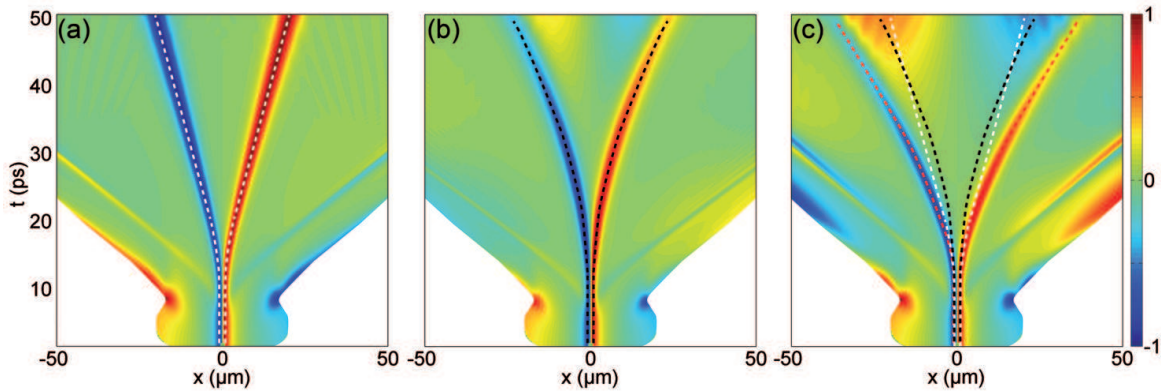}
\centering
\caption{Contributions to the HS motion. (a) $\alpha_2=-0.2 \alpha_1$ and $\hbar\Omega_x=0$, (b) $\alpha_2=0 \alpha_1$ and $\hbar\Omega_x=100$ $\mu$eV and (c) $\alpha_2=-0.2\alpha_1$ and $\hbar\Omega_x=100$ $\mu$eV. The dashed lines stand for the trajectories in the configuration of the panel (a) (white) and (b) (black) to be compared to the red line.}
\label{fig4}
\end{figure}

To underline the contributions to the solitons dissociation, we have numerically implemented a realistic configuration using GL beams in each component with winding numbers $(l_+,l_-)=(+1,+1)$ and $\phi_0=0$. We show in Fig.\ref{fig4} the HS propagation separating the different contributions in a situation where there is no rolling effect. In the panel (a), we show the $\alpha_2$ impact on the separation, with $\Omega_x=0$: we observe the linear trajectories (similarly to Fig.\ref{fig2}(c)) after the HS are released from the pulsed pump spot, no acceleration is observed. In the panel (b) only $\vec{\Omega}_{LT}=\Omega_x \vec{u}_x$ is present ($\alpha_2=0$), trajectories become parabolic up to the limiting speed, which is a clear signature for the constant monopole acceleration (non-relativistic limit). In the panel (c), we show the combined effect of both $\alpha_2$ and $\vec{\Omega}_{LT}$ together with the traces of the solitons trajectories (dashed lines) from the panels (a) and (b), in that case the acceleration is emphasized by the initial  repulsion. The rolling effect can either assist the separation or block it, depending on both the winding number imposed and the lateral shift direction.

We have seen that the splitting of a linearly polarized beam into its circularly polarized components allows an independent selection of the winding number of the half-solitons in each component and to fix their initial separation. However, even this step can be avoided, in order to further simplify the experimental configuration. Indeed, making the entire beam scatter on a hologram will impose the same winding number $l_+=l_-$ to both condensate components, and the two components will this time be perfectly superimposed on the sample. In that case, the separation will not occur, unless the symmetry between the $\sigma_+$ and $\sigma_-$ components is broken by some means. Although noise will always be present in the system and allow the separation to occur, it might induce (in some realizations) a symmetry breaking on time scales larger than the polariton lifetime. Moreover, the $\sigma_+$ and $\sigma_-$ solitons will be separated randomly (e.g. $\sigma_-$ going to left and $\sigma_+$ going to the right in one realization and the opposite behavior in the next one). These two points, that we have checked numerically (not shown), are clearly harmful to the reproductibility of the effect, and therefore, to the creation of a spin current. One should also bear in mind that a beam prepared to be linearly polarized might carry a small ellipticity. The ensuing density imbalance between the two condensate components will lead to the formation of HSs with slightly different healing lengths and therefore different effective masses. The rolling effect (which will occur in the same direction for both HSs) or possibly the wedge naturally present in microcavities (or the small gradient of the wire width), will induce an effective mass-dependent motion providing the separation. We have simulated this configuration with no initial separation, $(l_+,l_-)=(+1,+1)$ and a 1\% ellipticity of the input beam. We show in Fig.\ref{fig6} the separation obtained from the rolling effect shifting the beam by $\Delta y=+0.5$ $\mu$m [panel(a)] and from a wedge in the sample producing linear ramp potential of slope $10$ $\mu$eV/$\mu$m [panel(b)]. In that latter case the HSs are accelerated by the force they undergo not from the side of the magnetic field, but directly from the potential ($\hbar \Omega_x=0$ here).

\begin{figure}[ht]
\includegraphics[width=0.66\linewidth]{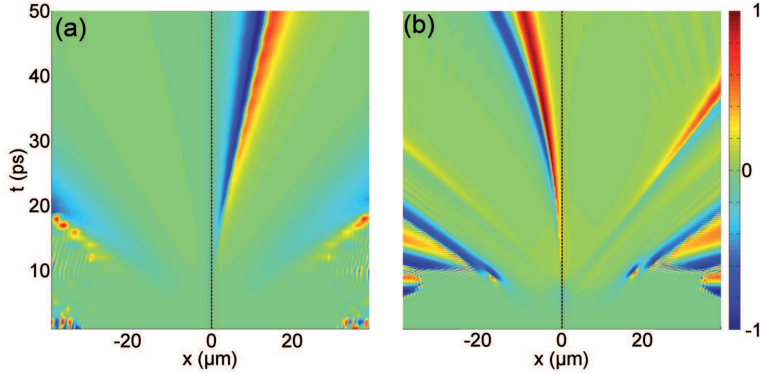}
\centering
\caption{Separation of perfectly overlapping HSs thanks (a) to the rolling effect induced by a $\Delta y=0.5$ $\mu$m shift and (b) to a wedge producing a potential ramp of slope $10$ $\mu$eV/$\mu$m. For both configurations $\alpha_2=-0.2 \alpha_1$, $\hbar \Omega_x=0$ and $l_+=l_-=+1$.}
\label{fig5}
\end{figure}

In conclusion, polaritons condensates in quasi-1D microwires constitute an ideal system for the creation of magnetic currents based on half-solitons thanks to their easy nucleation, natural separation and acceleration. The half-light component in the polariton wavefunction allows the particles to travel at high velocities and therefore the analogues of magnetic monopoles that are half solitons can travel at the speeds close to the speed of light. They constitute extremely promising entities for the fabrication of a new class of high speed spin-optronic devices. While the propagation of these magnetic charges in wires is technologically attractive, two dimensional systems (planar microcavities) allow even more freedom and fundamental richness encompassed in the angular momentum of vortices as we will discuss in the following section.

\section{Half-vortices separation}
Integer vortices in "scalar" polariton condensates (that is, circulary polarized, without significant coupling between the two components) have already been demonstrated experimentally in the optical parametric oscillator (OPO) configuration \cite{Whittaker,GLBeam}. These vortices have been shown to be relatively persistent, remaining in the ground state of the OPO for long times, much longer than polariton lifetime. Another configuration, where the integer and half-integer vortices occur due to the persistent flows in the polariton condensate under non-resonant pumping, has also been studied experimentally \cite{LagVort,LagHV}.

In this work we are not particularly interested in demonstrating the self-sustained coherence of polariton condensate maintained by the OPO. We propose a configuration which seems to be the simplest from the point of view of experimental realization, in order to observe controlled separation of half-vortices. Using the Gauss-Laguerre beam we create an integer vortex, which then evolves freely under the effect of spin-dependent polariton-polariton interactions, constant effective in-plane magnetic field, and $\vec{k}$-dependent TE-TM splitting. We demonstrate numerically the separation of a vortex $(l_+,l_-)=(+1,+1)$, with both components rotating in the same direction, in a weak static in-plane field of 10 $\mu$eV. Such integer vortex is likely to form in case of non-resonant pumping, due to persistent flows in the polariton condensate in a disorder landscape. The separation into half-vortices occurs for $\phi_0=\pi$, while for $\phi_0=0$ the two vortices remain coupled with each other. Indeed, the half-vortices of opposite charges form a dipole, and the interaction between these charges induced by the field increases with the applied field. This is why, depending on the initial winding numbers and on the pseudospin texture defined by $\phi_0$, one can sometimes observe the stabilization of the dipole length at an equilibrium value. Figure 7 shows the initial pseudospin texture (a) and the comparison between the situations without in-plane magnetic field (b) and with magnetic field (c). In the case (b) the small separation is due to the $\alpha_2$-induced repulsion. In the panel (c) the half-vortices are dissociated and accelerated in opposite directions by the constant in-plane magnetic field, although its value is relatively small.

The parameters of the Gauss-Laguerre beam have to be chosen carefully, in order to minimize the non-desired effects such as the dynamic formation of solitons on the outer horizon of the density profile. At the same time, the size of the minimum in the center should be in agreement with the expected healing length. In a word, one should be as close to the perfect initial condition of a flat infinite condensate with a vortex in the center, as possible.

\begin{figure}[h]
\includegraphics[width=0.99\linewidth]{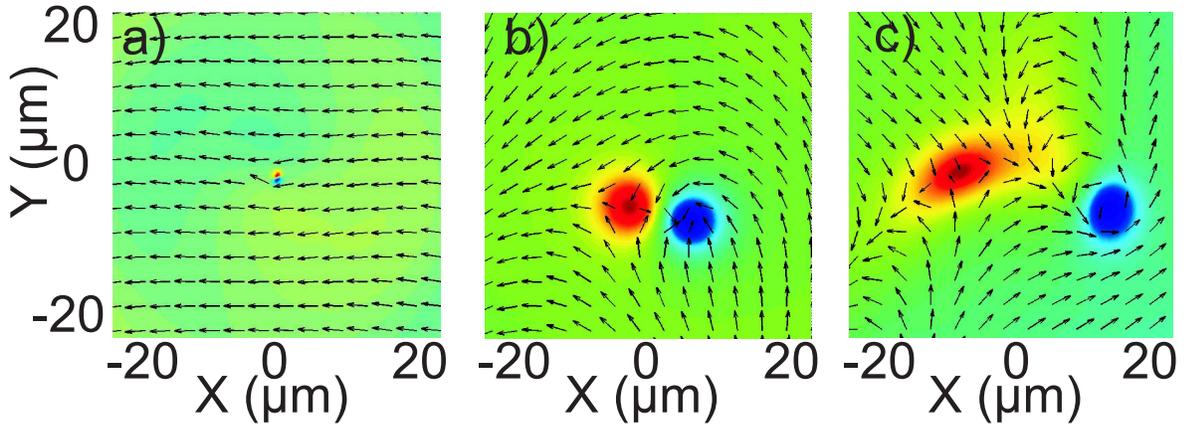}
\centering
\caption {a) Initial pseudospin configuration for an integer vortex (+1,+1); b) Dissociation into two HVs under the effect of $\alpha_2$-induced repulsion only; c) Dissociation and acceleration of the two HVs under the effect of magnetic field $\Omega_{x}$. Panels (b) and (c) correspond to 50 ps after the pulse.}
\label{fig6}
\end{figure}

However, peculiar undesired effects can arise if a half-vortex becomes trapped in the spatial density inhomogeneity created by the rotation of the pseudospin of the other half-vortex, which is not cylindrically symmetric: in some directions the pseudospin is initially aligned with the field and does not rotate, while in the other directions the pseudospin is not aligned with the field and is bound to gain a nonzero $Z$-projection, which can be a barrier or a trap for the other half-vortex.

When the half-vortices are sufficiently far from each other (that is, when the distance between the vortices $d\gg \xi$), each of them can be considered separately, as a half-vortex composed of a vortex in one component and a homogeneous background in the other component. Each of them is therefore subject to a force acting from the effective magnetic field, and we can expect them to accelerate freely. The pseudospin texture of a half-integer vortex can be either convergent or divergent, looking similar to that of a point charge, as it is the case in 1D, but there are two important differences between the 1D and the 2D case. Indeed, in 1D the constant pseudospin field far from the vortex core is exactly the solution of the Maxwell's equation $\vec{\nabla}\cdot  \vec{S}=\delta(x)$, whereas in 2D the pseudospin texture does not depend on the distance from the core (as well as in 1D), while the solution of the Maxwell's equation in this case should be decaying: the field of a charge depends on the distance from it. The second important difference resides in the presence of the relative phase $\phi_0$, which influences the more complicated textures of the half-vortex and its propagation direction. Various cases and the corresponding textures have been considered in Ref.\cite{SolnyshkovMonopoles}. Moreover, in the 2D case there is a long-distance interaction between the vortices lying in the same component (which is not the case for solitons in 1D), and even between vortices in different components, when a magnetic field is applied. The field creates a transfer of particles between the two components, and since the particles have nonzero propagation velocities even far from the vortex core, this creates a long-distance interaction force, absent in 1D. For the $\vec k$-dependent TE-TM splitting in planar microcavities, this interaction has been considered in Ref.\cite{Toledo}.

\begin{figure}[ht]
\centering
\includegraphics[width=0.75\linewidth]{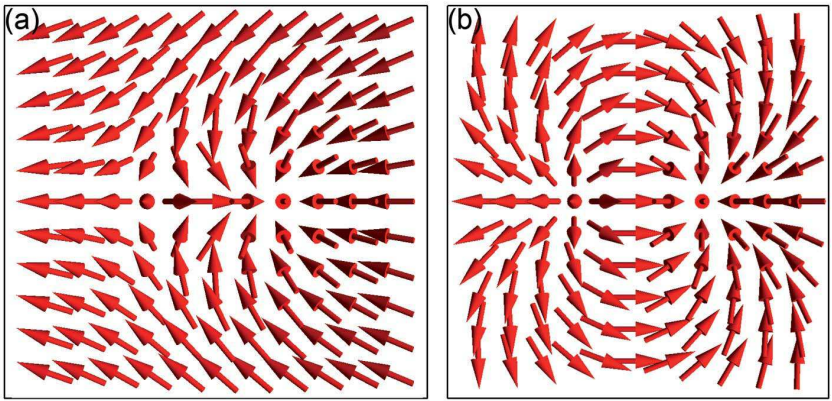}
\caption {Pseudospin textures of (a) phase vortex $(l_+,l_-,\phi_0)=(+1,+1,\pi)$ and (b) a polarization vortex $(l_+,l_-,\phi_0)=(1,-1,\pi)$, dissociated in their associated half-vortices.}
\label{HVTextures}
\end{figure}

In the ideal case, the condensate wavefunction containing a half-vortex in both components separated by $d$ reads
\begin{equation}\label{HVWF}
\vec \psi   = \left( \begin{array}{l}
{\psi _ + }\\
{\psi _ - }
\end{array} \right) = \left( \begin{array}{l}
\sqrt {{n_ + }} {e^{i{l_ + }\phi_+ }}{e^{i{\phi _0}}}\\
\sqrt {{n_ - }} {e^{i{l_ - }\phi_- }}
\end{array} \right)
\end{equation}
The radial functions take the approximated form \cite{BEC} $n_\pm(r_\pm,\phi_\pm)=r_\pm^2/(r_\pm^2+2)$, where $r_\pm=\sqrt{(x\mp d/2)^2+(y\mp d/2)^2}$ and $\phi_\pm=\arctan[(y \mp d/2)/(x \mp d/2)]$. The components of $\vec S$ are easily found using the Eqs.(6). To give the complete picture, we show in Fig.\ref{HVTextures} the pseudospin textures for two half-vortex pairs using the sets of parameters $(l_+,l_-,\phi_0)=(+1,+1,\pi)$ and $(l_+,l_-,\phi_0)=(1,-1,\pi)$ which constitute a phase and a polarization vortex respectively when the half-vortices overlap.

In conclusion, half-vortices embody the 2D candidates for the monopole analogy possessing a divergent pseudospin field (depending on $\phi_0$). They accelerate under the action of the effective field and are characterized as well by a $\phi_0$-dependent propagation direction \cite{SolnyshkovMonopoles} at fixed orientation of the effective field. They constitute the building blocks for potential 2D-magnetic circuits.

\section{Conclusions}
We have shown that the half-integer topological defects in spinor polariton condensates not only possess a divergent or convergent pseudospin texture similar to the textures of point charges in 1D and 2D, but also behave as magnetic charges in presence of effective magnetic fields, accelerating along them. We have proposed a simple configuration for experimental creation, dissociation, and acceleration of half-integer topological defects.
~~\\
~~\\
\emph{Acknowledgments}\\
We would like to thank J. Bloch, A. Amo and I. A. Shelykh for fruitful discussions. We acknowledge the support of FP7 ITN "Spin-Optronics" (237252), ANR "Quandyde", and IRSES "POLAPHEN" (246912) projects.

\end{document}